\documentstyle[12pt,epsf]{article}
\def\ZZ{\hbox{$Z$}
        \hbox{\kern -6pt  Z}}
\input{psfig}

\newcommand{\nc}{\newcommand}
\def\frac#1#2{{\textstyle {#1 \over #2}}}

\nc{\beq}{\begin{equation}}
\nc{\eeq}{\end{equation}}
\nc{\bea}{\begin{eqnarray}}
\nc{\eea}{\end{eqnarray}}
\nc{\lsim}{\begin{array}{c}\,\sim\vspace{-21pt}\\< \end{array}}
\nc{\gsim}{\begin{array}{c}\sim\vspace{-21pt}\\> \end{array}}
\nc{\eps}{\epsilon}
\nc{\s}{\sigma}
\nc{\veps}{\varepsilon}
\begin{document}
\begin{titlepage}

\begin{center} 

\vskip .5 in
{\large \bf 
Coupled critical Models: Application to Ising-Potts Models}

\vskip .6 in
 
  {\bf  P.  Simon}
   \vskip 0.3 cm
 {\it   Laboratoire de Physique Th\'eorique et Hautes Energies 
}\footnote{Laboratoire associ\'e No. 280 au CNRS}\\ 
 {\it  Universit\'es Denis Diderot Paris VII et Pierre et Marie Curie Paris VI}\\
{\it  2 pl. Jussieu, 75251 Paris cedex 05 }\\
\vspace*{.3cm}
{\it simon@lpthe.jussieu.fr}
  \vskip  1cm  
\end{center}
\vskip .5 in
\begin{abstract}
We discuss the critical behaviour of $2D$ Ising and $q-$states Potts models 
coupled by their energy density. We found  new tricritical points.
The procedure employed is the 
renormalisation approach of the perturbations series around conformal field 
theories representing pure models as already used for disordered spins models. This analysis could be useful to understand the physics of coupled critical models like the fully frustrated XY model.

\end{abstract}

\vskip 1cm
\noindent
PACS numbers: 64.60.Ak, 64.60.Fr, 05.70.Jk, 75.10.Hk\\
Keywords: critical points, perturbed conformal field theory, coupled minimal models.
\end{titlepage}%%%%%%%%%%%%%%%%%%%%%%%%%%%%%%%%%%%%%%%%%%%%%%%%%%%%%%%%%%%%%%%
\renewcommand{\thepage}{\arabic{page}}
\setcounter{page}{1}

\pagestyle{plain}
\pagenumbering{arabic} 
\vskip .5 truecm 
The critical behaviour of the two dimensional XY-Ising model, consisting of
XY and Ising models coupled though their energy densities has been the subject 
of an amount of work \cite{kos1,kos2}. The model is expected to include in its 
range of parameters a class of models with $U(1)\times Z_2$ symmetry as for 
example
the $2D$ fully frustrated XY model \cite{ffxy}, the $2D$ arrays of Josephson 
junctions \cite{joseph}, the $2D$ ANNNXY model \cite{sim} and so on. The phase
diagram of the XY-Ising model is interesting in its own right because of some 
unsual critical behaviour \cite{kos2}. Indeed, there is a plane in the space of 
parameters with a continuous critical line with simultaneous XY and Ising 
ordering which joins a branch point (where the XY and Ising transitions 
separate) and a tricritical point. Along this critical line, critical 
exponents associated to Ising-like order parameter were found to differ from 
pure Ising models. The central charge appears to increase continuously from 
$c\sim 1.5$ to $c\sim 2$. Nevertheless, some strong finite size effects have 
been shown and  therefore the value $c=1.5$ is finally not excluded 
\cite{kos2}. Hence, two scenarii are still possible, either the existence of a 
new universality class or simply the superposition of Ising and XY transitions 
with important finite size effects. \\Whatever the different scenarii, it seems important to understand in a more general context coupled crtical models. In a first step in this direction, we wonder in these letter whether such puzzling
behaviour could also be encountered in systems with $Z(2)\times 
Z(N)$ symmetries or more generally with $Z(2)\times Z(N)^n$. Namely, could the 
mixing of two different critical models by their density energy create new 
critical behaviour ? \\
 When $N=2$, the 
model consists of the n colored 
Ashkin-Teller model. This model is integrable when $n=2$ \cite{baxter}. For 
$n>2$  the transition is either first or second order  depending on the sign of 
the energy-energy coupling term \cite{teller}.
In the rest of the letter, we especially focus upon the case $N=3$ (the case 
$N=4$ will be a little discussed) because the $3-$states 
Potts model belongs to the unitary minimal serie.\\
Conformal field theory seems to be the best tool to describe the critical behaviour of such $2D$ critical models. Moreover,
perturbed conformal field theory has been recently a useful tool to investigate 
the 
critical properties of the two dimensional Ising and Potts models with quenched 
random bounds \cite{dotsen1,ludwig,cardy,pujol}. In this letter, we used the 
same 
formalism 
to investigate the critical behaviour of two pure models (Ising and $q-$states 
Potts models) where we add some energy-energy coupling terms. 
Such an analysis will also be useful to characterize new simple fixed points,
which could serve as tests for pertubation conformal field theory.
\\

Let us  consider the general case of $m$ critical Ising models and $n$ critical 
Potts models which are 
coupled by their energy density. 
The Hamiltonian has the following form
\beq
\label{ham}
H=\sum\limits_{a=1}^m H_{Ising}^a+\sum\limits_{b=m+1}^{m+n} H_{potts}^b- 
\sum\limits_{c,d} g_{cd} \int d^2 z~\varepsilon_c(z)\varepsilon_d(z)~,
\eeq
where  $c,d$ belongs to $[1..m+n]$. $H_{Ising}^a$ and $H_{potts}^b$ represent 
respectively pure Ising models and Potts models at their critical temperatures. $\varepsilon_c$ is the energy operator of an Ising or Potts model depending on the value of $c$.
 We have a $(m+n)\times 
(m+n )$ matrix of coupling constant. We will study the influence of this last term 
 on pure 
models in perturbation.

The partition function can hence be written as
\beq
\label{z}
Z = \prod_{a=1}^m Tr_{a,s_{i,a}} \prod_{b=m+1}^{m+n} 
Tr_{b,s_{i,b}}e^{-\displaystyle\sum_{a=1}^m
S_{0,a} -\sum_{b=m+1}^{m+n}
S_{0,b}+  \int \sum\limits_{c\ne d}g_{cd}\displaystyle
\varepsilon_c(z)\varepsilon_d(z) d^2z ~.
}
\eeq
In our notation, $a$ corresponds to Ising models, $b$ to Potts models, and $ s_{i,c}$ is the spin operator of the $c^{th}$ model.
A correlation
function $<O(0)O(R)>$, where $O$ is some local operator, is expanded 
perturbatively like:
$$
<O(0)O(R)> = <O(0)O(R)>_0+<S_IO(0)O(R)>_0+{1\over2}<S_I^2O(0)O(R)>_0+\cdots
$$
where $<>_0$ means the expectation value taken with respect to $S_0=\sum\limits_{a=1}^m
S_{0,a} +\sum\limits_{b=m+1}^{m+n}
S_{0,b}$ and
\beq
S_I =
\int H_I(z) d^2z =  \int \displaystyle\sum_{c\not=d}g_{cd}
\varepsilon_c(z)\varepsilon_d(z) d^2z~.
\eeq
The calculation of correlation functions can be performed with the Coulomb-gas 
representation \cite{dotsen2}. The central charge is written as $c={1\over 
2}+\epsilon$, where $\epsilon$ will be used as a short distance regulator for 
the integrals involved in correlation functions calculations. In addition, we also use an I.R. cut -off $r$. Then the limit 
$\eps\to 0$ corresponds to the Ising model while the Potts model is obtained 
for some finite value of $\eps$.
The central charge $c$ will be
characterized in the following by the parameter $\alpha^2_+ = {2p\over2p-1}
={4\over 3} + \epsilon $ with
\bea
c=1-24\alpha_0^2~~&;&~~\alpha_{\pm}=\alpha_0\pm\sqrt{\alpha_0^2+1};\\
\alpha_+\alpha_-&=&-1 .\nonumber
\eea
For the pure Ising model   $\alpha_+^2={4\over 3}$ and $c={1\over 2}$ while for 
the $3-$states Potts model  $\alpha_+^2={6\over 5}$, $c={4\over 5}$ and 
$\eps=-{2\over 15}$. The $4-$states Potts model can be considered as the limiting case of this  perturbative scheme \cite{dotsen1}, it has c=1 and $\eps=-{1\over 3}$. The vertex operators are defined by 
$V_{nm}(x)=e^{i\alpha_{nm}\Phi(x)}$ where $\Phi(x)$ is a free scalar field and 
the $\alpha_{nm}$ are defined by
\beq
\alpha_{nm} = {1\over 2}(1-n)\alpha_-+{1\over 2}(1-m)\alpha_+~.
\eeq
The conformal dimension of the operator $V_{nm}(x)$ is 
$\Delta_{nm}=-\alpha_{\overline{nm}}\alpha_{nm}$ with
\beq
\alpha_{\overline{nm}}=2\alpha_0-\alpha_{nm}={1\over 2}(1+n)\alpha_-+{1\over 
2}(1+m)\alpha_+~.
\eeq
The spin field $\sigma$ can be represented by the vertex operator $V_{p,p-1}$ 
whereas $V_{1,2}$ corresponds to the energy operator $\veps$. Note that in the 
Ising case the $\sigma$ operator could also be represented by the $V_{2,1}$. 
When we have only one coupling constant $g_0$, 
its renormalisation is determined
directly by a perturbative computation. $g$ is also
given by the O.A. producing
\beq
g_0 \int \displaystyle\varepsilon_a(z)\varepsilon_b(z) d^2z
+ {1\over 2} \left(g_0 \int \displaystyle
\varepsilon_a(z)\varepsilon_b(z) d^2z\right)^2 +\cdots= g\int 
\displaystyle\varepsilon_a(z)\varepsilon_b(z) d^2z~,
\eeq
with $g=g_0 + A_2 g_0^2  + \cdots $ where $A_2$ comes from the contraction
\beq
{1\over 2} \int \displaystyle 
\varepsilon_a(z)\varepsilon_b(z) d^2z \int \displaystyle 
\varepsilon_c(z)\varepsilon_d(z) d^2z = A_2 \int
\displaystyle\  \varepsilon_a(z)\varepsilon_b(z) d^2z 
\eeq
\noindent
This procedure generalizes in a straighforward way to our case. If we suppose
that $g_{cd}=g_1$ for $c,d \in [1,m]$, $g_{cd}=g_2$ for $c,d \in [m+1,m+n]$ and
finally  $g_{cd}=g_{12}$ for $c\in [1..m], d\in[m+1,m+n]$ or vice versa, we 
obtain the RG equations associated to these three coupling constant (at second 
order)
\bea
\label{systeme}
\beta_1={d g_1\over d(ln l)}&=& 4\pi(m-2)~g_1^2+4\pi n~ g_{12}^2~+O(g^3),
\nonumber\\
\beta_2={d g_2\over d(ln l)}&=& ~-3\eps g_2+4\pi(n-2)~g_2^2+4\pi m 
~g_{12}^2~+O(g^3),\\
\beta_{12}={d g_{12}\over d(ln l)}&=& -3{\eps\over 2} 
g_{12}+4\pi(m-1)~g_1g_{12}+4\pi (n-1)~g_2g_{12}~+O(g^3).\nonumber
\eea
We have introduced a UV regulator $\eps$. 
In order to recover the case of Ising coupled to Potts models, we will take at the end 
the limit $\eps\to -{2\over 15}$ for the $3-$states Potts model and  $\eps\to -{1\over 3}$
for the $4-$states Potts model.
Let us study the fixed point structure in the general case.\\
When $m>1,n\ge1$ we have only (at this order) two decoupled structures   defined 
by 
$g_{12}^*=0$. It corresponds in the infrared limit to the $m-$colored Ashkin 
Teller model \cite{baxter,teller} plus a $n-$colored potts model. The critical 
behaviour of the latter is not properly established: for $n=2$, the model turns 
out to be integrable presenting a mass generation indicating a first order 
transition\cite{vays} whereas for $n>2$ no exact solution exists. Nevertheless 
a perturbation analysis to third order tends to show a first order transition 
for $g_{2}>0$ and a new non-trivial infrared fixed point for $g_{2}<0$ 
\cite{pujol}. At this level, we
have not new non trivial fixed points mixing both models. \\
If we consider now the peculiar case $m=n=1$. The coupling constant $g_{12}$ 
always flows in a strong coupling regime indicating probably a mass generation.
Consequently,  when one superposes a critical Ising model to a critical $3$ or $4$ states potts model, no new critical behaviour is found  pertubatively when coupling both model by their energy density. Of course, new non-perturbative fixed points cannot be excluded by this method. Therefore, It suggests (perturbatively) that no line with simultaneous disordering of the 
Ising and Potts order parameter survives in the phase diagram of this model contrary to those of the FFXY model. 
Let us now treat the most interesting case $m=1,n>1$ as mentioned in the 
introduction.
Six fixed points are found:
\bea
\label{ptfix1}
&&g_1^*=g_2^*=g_{12}^*=0\\
\label{ptfix2}
&&g_1^*=g_{12}^*=0;~g_2^*={3\eps\over 4\pi(n-2)}~~ if~~ n>2\\
\label{ptfix3}
&&g_1^*={-3\eps\tau_1 n\over 8\pi(n-1)};~g_2^*={3\eps\over 
8\pi(n-1)};~g_{12}^*={-3\eps\tau_2\sqrt{n}\over 8\pi(n-1)}~~ if~~ n>1
\eea
with $\tau_1,\tau_2=\pm1$.
There are now four new non trivial fixed point (\ref{ptfix3}). As usual, we can 
study the stability of each of these fixed point by re-expressing 
(\ref{systeme}) around the above fixed points ($g_i=g^*_i+\delta g_i$) and keeping only the smallest 
order in $\eps$. This will give us a linear system $\delta\dot{g}=A~\delta g$, with $\delta g$ a three component vector and $A$ a real $3\times3$ matrix.
%$$
%\left( {\delta \dot{g} \atop \delta  \dot{\Delta}}\right) = A
%\left( {\delta g \atop \delta  \Delta}\right)  
%$$
The stability of these fixed points can be obtained by calculating the eigenvalues of $A$
in each of these cases. In this way, it is easy to see that (\ref{ptfix1}) is unstable, and that (\ref{ptfix2}) is stable in the $(g_2,g_{12})$ plane and marginal in the direction $g_1$. It  corresponds to the infrared fixed point of $n$ coupled Potts models \cite{pujol}. Concerning the other four fixed points (\ref{ptfix3}), two are unstable in two directions (when $\tau_1=-1$) and the order two ($\tau_1=+1$) are hyperbolic fixed points, namely stable in two directions. They constitute new tricritical points. The stable plane is defined by the vectors $(1,0,0);(0,{\tau_2\sqrt{2n(n-1)+1}-1\over \sqrt{n}(n-1)},1)$ in the $(g_1,g_2,g_{12})$ space. A projection of the flow in the
$(g_{12},g_2)$ plane has been given in Figure 1 for the case $n=3$. The flow is symetric along the $g_2$ axis. If initially $g_2^0>0$ then the flow is always driven in a strong coupling regime indicating mass generation as for the simple $n$-colored Potts model \cite{pujol}. Nevertheless, if $g_2^0<0$ and small, the trajectory will be first attracted by one of the tricritical point (depending on the sign of $g_{12}^0$) and then flow toward the stable fixed point of the $n$-colored Potts model.   For the case $n=2$, the fixed point on the $g_2$ axis does not exist, so we always flow in the strong coupling regime except when we are on the stable plane in the $(g_1,g_2,g_{12})$ space. It is a reminescence of the exactly integrable $2$-colored Potts model which is always massive \cite{vays}.
\\\noindent
We want to characterize more accurately these new tricritical points. In this perspective, we will compute the renormalisation of the critical exponents associated to the spin-spin
correlation function. Hence, we need to calculate the multiplicative functions $Z_{\sigma_1}$ for the Ising spins and $Z_{\sigma_2}$ for Potts spins. A convenient way to define $Z_{\sigma}$ is to add to the general Hamiltonian (\ref{ham}) source terms for the spin operators as $h_1^0\int\sigma_1(z) d^2z+h_2^0\int\sum\limits_{a=1}^n\sigma_2(z)d^2z$. Hence, the multiplicative functions $Z_{\sigma}$  are defined by 
$$  
h_1\int\sigma_1(z) d^2z+h_2\int\sum\limits_{a=1}^n\sigma_2(z)d^2(z)
=h_1^0 Z_{\sigma_1}\int\sigma_1(z)  d^2z+h_2^0Z_{\sigma_2}\int\sum\limits_{a=1}^n\sigma_2(z)d^2z~.
$$
As for the coupling constant $g_i$, the functions $Z_{\sigma}$  can be computed in perturbation. The first contribution happens to second order. For the details of calculations, we have refered to  \cite{dotsen1} where rather similar technics have been used. Therefore, the $\sigma$ fields get renormalized as
\bea
\sigma_1&\to&\s_1(1+A_2 (g_{12}^0)^2+A_3(g_{12}^0)^2g_2^0+\cdots)\equiv Z_{\s_1}\s_1\\
\sigma_2&\to&\s_2(1+B_2 (g_{2}^0)^2+C_2 (g_{12}^0)^2+B_3(g_{2}^0)^3+ C_3(g_{12}^0)^2g_2^0+\cdots)\equiv Z_{\s_2}\s_2
\eea
with 
\bea
A_2&=&{\pi^2 n\over 2}r^{-3\eps}\\
A_3&=&-8n(n-1)\pi^3{r^{-6\eps}\over 6\eps}\\
 B_2&=& (n-1)\pi^2{r^{-6\eps}\over 2}\left[1+2{\Gamma^2(-{2\over 3})\Gamma^2({1\over 6})
\over \Gamma^2(-{1\over 3})\Gamma^2(-{1\over 6})}\right]\\
C_2&=&\pi^2{r^{-3\eps}\over 2}\left[1+4{\Gamma^2(-{2\over 3})\Gamma^2({1\over 6})
\over \Gamma^2(-{1\over 3})\Gamma^2(-{1\over 6})}\right]\\
B_3&=&-12\pi^3(n-1)(n-2){r^{-9\eps}\over 9\eps}\left[1+{4\over 3}{\Gamma^2(-{2\over 3})\Gamma^2({1\over 6})
\over \Gamma^2(-{1\over 3})\Gamma^2(-{1\over 6})}\right]\\
C_3&=&-8\pi^3(n-1)({r^{-6\eps}\over6\eps})\left[1+2{\Gamma^2(-{2\over 3})\Gamma^2({1\over 6})\over \Gamma^2(-{1\over 3})\Gamma^2(-{1\over 6})}\right]
\eea
In fact, to obtain such results, we have to conserve two  regulators $\eps_1$ for Ising operators,$\eps_2=\eps$ for Potts operators,
first to follow the dimensions of expressions and especially because some  divergent terms (in ${1\over \eps_1}$) appears at one and two loops. These terms disappear when we replace the  bare coupling constants by renormalised ones assuring the renormalisability of the theory. They have not been included for readability.\\Therefore,
using the remormalisation group equations for $g_i$ (\ref{systeme}), the equations for $Z_{\s_i}$ becomes
\bea
{d\log(Z_{\s_1})\over d\log(r)} &=&-3ng_{12}^2(r){\pi^2\over 2}\eps+4n(n-1)\pi^3g_{12}^2(r)g_2(r)\\
{d\log(Z_{\s_2})\over d\log(r)} &=&-3(n-1)g_2^2\pi^2\eps\left[1+2{\cal R}\right]-3g_{12}^2{\pi^2\over 2}\eps\left[1+4{\cal R}\right]\nonumber\\&&+4(n-1)(n-2)\pi^3g_2^3(r)+4(n-1)\pi^3g_{12}^2(r)g_2(r)
\eea
where we have defined $${\cal R}={\Gamma^2(-{2\over 3})\Gamma^2({1\over 6})
\over \Gamma^2(-{1\over 3})\Gamma^2(-{1\over 6})}$$
The Callan-Symanzik equation gives the form of the correlations functions between  spins $\s$. For one coupling constant $g$, we have
\beq
\label{exp}
<\sigma(0)\sigma(sL)>_{a,g_0} =
e^{2\int\limits_{g_0}^{g(s)}{\gamma_{\s}(g)\over \beta(g)}
dg}s^{-2\Delta_{\sigma}} <\sigma(0)\sigma(L)>_{r,g(s)}
\eeq
where we used the notation :
$
{dln(Z_\sigma) \over dln(r)} = \gamma_{\s}(g)
$
; $ g(a)= g_0 $ and $g(s)$ is defined by
$\int\limits_{g_0}^{g(s)} \beta(g) dg = ln(s)$; $r=sa$, and $a$ is a
lattice cut-off scale. $L$ is an arbitrary scale which can be fixed to one lattice spacing $a$ of a true statistical model. The dependence in $s$ of the term $ <\sigma(0)\sigma(L)>_{r,g(s)}$ is thus negligible and can be considered as a constant. We want to compute the corrections to the critical exponents close at the infrared tricritical points defined in (\ref{ptfix3}). So, the integral is dominated by the region $g\sim g^*$ and we have $\int\limits_{g_0}^{g(s)}{\gamma_{\s}(g)\over \beta(g)}
dg\sim\gamma_{s}(g^*)\log(s)$. We thus obtain
\beq
<\sigma(0) \sigma(s)>_{g_0} \sim
s^{-(2\Delta_{\sigma}-2\gamma_{\s}(g^*)) }
\eeq
Therefore, we can compute the corrections to critical exponents associated to spin operators $\s_i$ using the expressions of the tricritical points (\ref{ptfix3})
\bea
\gamma_{\s_1}(g_c)&=&-3n{\pi^2\over 2}\eps (g_{12}^*)^2+4n(n-1)\pi^3(g_{12}^*)^2g_2^* 
=0+O(\eps^4)\\
\gamma_{\s_2}(g_c)&=&-3(n-1)(g_2^*)^2\pi^2\eps\left[1+2{\cal R}\right]-3(g_{12}^*)^2{\pi^2\over 2}\eps\left[1+4{\cal R}\right]\nonumber\\&&+4(n-1)(n-2)\pi^3(g_2^*)^3+4(n-1)\pi^3(g_{12}^*)^2g_2^*\nonumber\\
&=&{-27n\eps^3\over 128(n-1)^2}-{27(2n-1)\eps^3\over 32(n-1)^2}{\cal R}+O(\eps^4),
\eea
with $\eps={-2\over 15}$ for the $3$-state Potts model and $\eps={-1\over 3}$ for the $4$-state Potts model.
Hence, we obtain the unavoided result, that there is no renormalisation of the critical exponent $\Delta_{\s_1}$ associated to the Ising spin at third order in $\eps$! It is diificult to give a particular interpretation to this result. Yet, the corrections to   $\Delta_{\s_2}$ are non zero and different from those of the infrared fixed point of n coupled Potts models \cite{pujol}. Therefore, these two tricritical points are new and non-trivial. We hope that these corrections to the critical exponents of the spin correlations functions could be useful to test numerically the results presented in this letter. Before concluding, a few remarks need to be done: Suppose we have a theory described by  one of the tricritical fixed points $T_i$, $i=1,2$  (see Figure 1), then a small perturbation can send the flow   to the fixed point $D$ characterized by decouple Potts and Ising models. The central charge at this point  $c_{D}$ is thus the sum of the central charges of each model. Therefore according to Zamolodchikov c-theorem \cite{zamolod}, the central charge at $T_i$ ($c_{T_i}$) verifies $c_{T_i}\ge c_{D}$. A similar scenario could be imagined for the FFXY model with a new tricritical point inducing  a strong cross-over regime (it would justify strong finite size effects and $c_{FFXY}>{3\over 2}$).\\

We have thus shown that the behaviour of coupled minimal models is far from trivial. Two new tricritical fixed points have been identified using perturbation around Ising model.
Maybe, a similar procedure could be investigated for Ising-XY models with another perturbation scheme. The study of these models with weak disorder is also interesting and wil be presented elsewhere in a longer paper where some details of the present paper could be found.

\vskip .5 in
{\bf Acknowledgements}\\
I would particularly like to thank Vl.S. Dotsenko for many useful suggestions and comments on this work.
Stimulating discussions with M. Picco. are also acknowledged.

\eject
%\begin{center}
%{\bf FIGURE CAPTIONS}
%\end{center}

\vskip 0.5 truecm
FIG.1 : The projection of the flow in the $(g_{12},g_2)$ plane for one Ising model coupled to three Potts models. Two tricritical points are found.

\begin{figure}
\epsfxsize=14cm
$$
\epsfbox{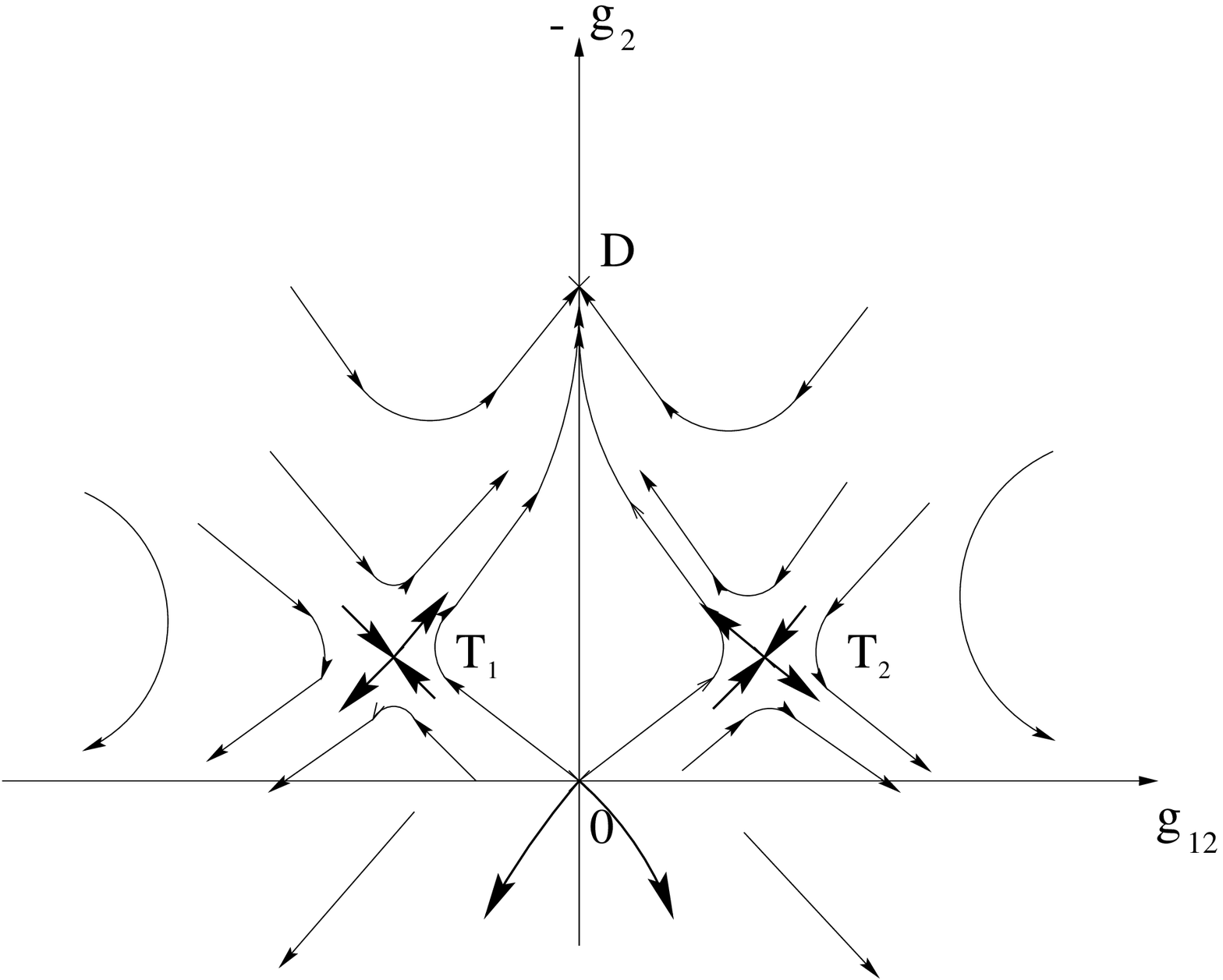}
$$

\end{figure}
\end{document}